\documentclass[twocolumn]{aastex631}

\usepackage{hyperref}
\usepackage{float}
\usepackage{threeparttable}

\usepackage{amsmath}
\usepackage{amssymb}
\usepackage{xspace}
\usepackage{xifthen}
\usepackage{eso-pic}

\newcommand{\Gaia}{{\it Gaia}\xspace}

\newcommand{\secref}[1]{Section~\ref{sec:#1}}

\newcommand{\tabref}[1]{Table~\ref{tab:#1}}
\newcommand{\figref}[1]{Figure~\ref{fig:#1}}
\newcommand{\eqnref}[1]{Equation~\eqref{eqn:#1}}

\mathchardef\mhyphen="2D
\newcommand{\roughly}{\ensuremath{ {\sim}\,} }
\newcommand{\code}[1]{\texttt{#1}\xspace}

\newcommand{\unit}[1]{\ensuremath{\mathrm{\,#1}}\xspace}
\newcommand{\uni}[1]{\ensuremath{\mathrm{#1}}\xspace}
\newcommand{\mac}{\unit{mag~arcsec^{-2}}}
\newcommand{\masy}{\unit{mas~yr^{-1}}}
\newcommand{\mas}{\unit{mas}}
\newcommand{\asec}{\uni{\arcsec}}
\newcommand{\amin}{\uni{\arcmin}}
\newcommand{\degree}{\ensuremath{{}^{\circ}}\xspace}
\newcommand{\magn}{\unit{mag}}
\newcommand{\Gyr}{\unit{Gyr}}
\newcommand{\kpc}{\unit{kpc}}

\newcommand{\bandvar}[2][]{%
  \ifthenelse{\isempty{#1}}{\var{#2}}{\var{#2\_#1}}%
}
\newcommand{\spreadmodel}[1][]{\bandvar[#1]{spread\_model}}
\newcommand{\magpsf}[1][]{\bandvar[#1]{mag\_psf}}
\newcommand{\errpsf}[1][]{\bandvar[#1]{errpsf\_world}}

\providecommand{\ra}{\ensuremath{\mathrm{R.A.}}}
\providecommand{\dec}{\ensuremath{\mathrm{decl.}}}
\providecommand{\mra}{\ensuremath{\alpha}}
\providecommand{\mdec}{\ensuremath{\delta}}
\providecommand{\pmra}{\ensuremath{\mu_{\mra\ast}}\xspace}
\providecommand{\pmdec}{\ensuremath{\mu_\mdec}\xspace}

\newcommand{\var}[1]{\ensuremath{\texttt{\MakeUppercase{#1}}}\xspace}
\newcommand{\SWarp}{\code{SWarp}}

\newcommand{\SExtractor}{\code{SExtractor}}
\newcommand{\sextractor}{\SExtractor}
\newcommand{\PSFEx}{\code{PSFEx}}

\newcommand{\astropy}{\code{Astropy}}
\newcommand{\modelsd}{\code{Fittable2DModel}}

\newcommand{\PMO}{\affiliation{Purple Mountain Observatory, Chinese Academy of Sciences, Nanjing 210023, China}}
\newcommand{\USTC}{\affiliation{School of Astronomy and Space Sciences, University of Science and Technology of China, Hefei 230026, China}}
\newcommand{\TDLI}{\affiliation{Tsung-Dao Lee Institute and State Key Laboratory of Dark Matter Physics, Shanghai Jiao Tong University, Shanghai, 201210, China;}}
\newcommand{\DAUSTC}{\affiliation{Department of Astronomy, University of Science and Technology of China, Hefei 230026, China}}
\newcommand{\IDSS}{\affiliation{Institute of Deep Space Sciences, Deep Space Exploration Laboratory, Hefei, 230026, China}}
\newcommand{\CASLab}{\affiliation{CAS Key Laboratory for Research in Galaxies and Cosmology, Department of Astronomy, University of Science and Technology of China, Hefei, 230026, China}}
\newcommand{\LabPDE}{\affiliation{State Key Laboratory of Particle Detection and Electronics, University of Science and Technology of China, Hefei 230026, China}}
\newcommand{\NOAO}{\affiliation{National Optical Astronomy Observatory (NSF's National Optical-Infrared Astronomy Research Laboratory), 950 N Cherry Ave, Tucson Arizona 85726, USA}}
\newcommand{\NAOJ}{\affiliation{National Astronomical Observatory of Japan, National Institutes of Natural Sciences, Tokyo 181-8588, Japan}}
\newcommand{\IOE}{\affiliation{Institute of Optics and Electronics, Chinese Academy of Sciences, Chengdu 610209, China}}

\shorttitle{Boötes III \& Draco}
\shortauthors{Yang et al.}

\begin{document}

\title{A Glimpse of Satellite Galaxies in the Milky Way with the 2.5-meter Wide Field Survey Telescope (WFST): Boötes III and Draco}

\author{Chao Yang}
\PMO
\USTC

\author{Zhizheng Pan}
\PMO
\USTC

\author{Min Fang}
\PMO
\USTC

\author{Xian~Zhong~Zheng}
\TDLI
\PMO

\author{Binyang Liu}
\PMO

\author{Guoliang Li}
\PMO
\USTC

\author{Tian-Rui~Sun}
\PMO

\author{Ji-An Jiang}
\DAUSTC
\USTC
\NAOJ

\author{Miaomiao Zhang}
\PMO

\author{Zhen Wan}
\DAUSTC
\USTC

\author{Shuang Liu}
\PMO
\USTC

\author{Han Qu}
\PMO
\USTC

\author{Ji Yang}
\PMO
\USTC

\author{Xu Kong}
\DAUSTC
\USTC
\IDSS

\author{Wenhao Liu}
\PMO
\USTC

\author{Yiping Shu}
\PMO
\USTC

\author{Jiang Chang}
\PMO
\USTC

\author{Tinggui Wang}
\DAUSTC
\USTC
\IDSS

\author{Lulu Fan}
\DAUSTC
\USTC
\IDSS

\author{Yongquan Xue}
\DAUSTC
\USTC

\author{Wentao Luo}
\IDSS

\author{Hongxin Zhang}
\DAUSTC
\CASLab

\author{Zheng Lou}
\PMO

\author{Haibin Zhao}
\PMO
\USTC

\author{Bin Li}
\PMO

\author{Hairen Wang}
\PMO

\author{Dazhi Yao}
\PMO

\author{Jian Wang}
\LabPDE
\IDSS

\author{Hongfei Zhang}
\LabPDE

\author{Feng Li}
\LabPDE

\author{Hao Liu}
\LabPDE

\author{Ming Liang}
\NOAO

\author{Jinlong Tang}
\IOE

\author{Yuheng Zhang}
\PMO
\USTC

\author{Man Qiao}
\PMO
\USTC

\author{Run Wen}
\PMO
\USTC

\author{Zongfei Lyu}
\PMO
\USTC

\correspondingauthor{Zhizheng Pan}
\email{panzz@pmo.ac.cn}

\begin{abstract}
We carry out deep imaging of the Milky Way satellite galaxies, Boötes III and Draco, with WFST as one pilot observing program to demonstrate the capability of WFST. Combining catalogs with PS1 DR2 and \Gaia DR3, we derive proper motions for candidate member stars in these two satellite galaxies over a 12-year time baseline, yielding uncertainties of $\roughly1.8\masy$ at 21\magn and $\roughly3.0\masy$ at 22\magn in the $r$ band. The proper motions derived from bright and faint stars are consistent, indicating no significant variation in proper motion across stellar luminosity as these galaxies undergo tidal interactions with the MW. Meanwhile, we suggest that Boötes III represents the bound remnant of the progenitor galaxy that gave rise to the Styx stream, as evidenced by its elongated density profile and overdensity in both spatial and kinematic space. This is the first paper to use WFST to measure the proper motions of faint stars in Milky Way satellite galaxies. More detailed analyses will be presented in forthcoming papers from the wide field survey (WFS) program.

\end{abstract}


\section{Introduction} 
\label{sec:intro}

Satellite galaxy serves as key component of the Milky Way (MW) dark matter halo \citep{Malhan2022}. As they orbit the MW, tidal forces gradually strip stars from their outer regions, stretching them into elongated streams \citep{Bonaca2025}. Since the deployment of \Gaia mission \citep{Gaia2016}, more accurate proper motion (PM) measurements for individual stars \citep[$\roughly0.6\masy$ at 20\magn in the $G$ band,][]{Lindegren2018,Lindegren2021} have been provided to estimate the PM and establish 6D phase diagrams (3D locations and 3D velocities) for the MW satellite galaxies and their stellar streams \citep{McConnachie2020,Lihefan2021}. Therefore, these satellite galaxies and stellar streams provide probes for measuring the gradient of the gravitational potential well \citep{Bonaca2014,McMillan2017} and constraining the mass distribution of the MW \citep{Li2017,Callingham2019,Slizewski2022}. 

Moreover, the satellite galaxies in the MW are predominantly dwarf galaxies, with the notable exceptions of the Large Magellanic Clouds and Small Magellanic Clouds. The shallow gravitational potential well of these dwarf satellite galaxies makes them easier to lose gas. This process inhabits star formation and metal enrichment in their evolutionary history. As a result, dwarf galaxies are primarily composed of old and metal-poor stars \citep{Jenkins2021,Fu2022,Fu2023,Lucchesi2024,Pan2025}. Additionally, observations from the Hubble Space Telescope (HST) have shown that they contain more low-mass stars than high-mass ones \citep{Geha2013,Gennaro2018,Filion2022}, and that their stellar populations are dominated by main-sequence stars \citep{Filion2020}. This implies that deeper imaging is more likely to resolve a higher fraction of faint member stars relative to bright ones. The spatial distribution of these faint stars reveals finer structural details of these galaxies \citep{Mu2018}. However, the limiting magnitude of \Gaia remains a challenge for estimating PMs of faint stars and ultral-faint satellite galaxies. 

Studies on the MW dwarf satellite galaxies can date back to 1938, when \cite{Shapley1938} discovered Sculptor, the first known dwarf satellite galaxy. Since then, more than 70 dwarf satellite galaxies in the MW have been discovered \citep{Harrington1950,Wilson1955,Cannon1977,Irwin1990,Ibata1994,Willma2005,Walsh2007,koposov2015,Drlica2015,Cerny2023b,Cerny2023c}. The identification of these systems generally relies on the spatial concentration of stellar distribution. Deep surveys, such as the Sloan Digital Sky Survey \citep[SDSS,][]{Gunn2006,Abazajian2009}, the Dark Energy Survey \citep[DES,][]{Flaugher2015,Abbott2021}, and the Hyper Suprime-Cam Subaru Strategic Program \citep[HSC-SSP,][]{Aihara2018,Aihara2022} have resolved more fiant stars with concentrated distribution, leading to the discovery of additional faint dwarf satellite galaxies. On the other hand, Enhancing the precision in distinguishing member stars from foreground stars and background galaxies also helps to refine the identification of these dwarf galaxies. 

Over the past two decades, the matched-filter method \citep{Walsh2009,Bressan2012}, which utilizes the color and magnitude information of stars to select an old, metal-poor stellar population, has been proven to be highly efficient in reducing contamination from foreground stars and background galaxies. However, wider and deeper surveys are time-consuming. In addition, stars may suffer differential reddening, particularly in the galactic disc, thus displaying complicated varying behaviour in colour-magnitude space. \cite{Torrealba2019} utilized RRL stars and PM data from \Gaia DR2 to discover an enormous satellite galaxy (Antlia 2), demonstrating the power of astrometric and light-curve information of stars to identify satellite galaxies, even in low galactic latitude regions.

Given a time baseline of at least 12 years from the Wide Field Survey Telescope (WFST) observations and the $3\pi$ Survey of the Panoramic Survey Telescope and Rapid Response System \citep[Pan-STARRS,][]{Onaka2012,Chambers2016}, it is possible for us to derive high-precision proper motion measurements for stars brighter than $r=22.5\magn$. Moreover, WFST is an imaging survey facility equipped with a 2.5-meter primary mirror and a mosaic CCD camera with 0.73 gigapixels on the primary focal plane \citep{Zhang2024}. Supported by automated Camera Control System \citep[CCS,][]{Geng2024}, Access Control System \citep[ACS,][]{Sun2024}, Telescope Control System (TCS), Observatory Control System \citep[OCS,][]{Zhu2024,Cao2024}, along with a wide field of view of $\roughly 6.5 \deg^{2}$, WFST is scheduled to complete a wide field survey (WFS) program covering $\roughly 8{,}000 \deg^{2}$ in the Northern Sky across four bands ($u$, $g$, $r$, and $i$) over a 6-year period \citep{Chen2024}. Since the WFS footprint is within the footprint of the $3\pi$ Survey, PMs for faint stars within this $8{,}000 \deg^{2}$ area can be derived immediately at the start of the WFS program. As one pilot observing program, our observations of Boötes III and Draco were carried out , serving as test observations for these two confirmed MW satellite galaxies \citep{Wang2023}. By combining data from \Gaia DR3, PS1 DR2, and our own catalogs, we aim to derive PMs for both bright and faint member stars, providing a more detailed characterization of the morphology and kinematic properties of these two satellite galaxies.

We organize our paper as follows: \secref{obs} introduces our observations; \secref{reduction} outlines the data reduction process, from raw images to final catalogs; \secref{selection} details the selection criteria for member stars in the two satellite galaxies; \secref{results} presents the main results, including the PMs of both galaxies, the tidal structure of Boötes III, and its potential connection to the Styx stellar stream; finally, \secref{sum} provides a summary and future outlook.

\section{Observation} 
\label{sec:obs}

Considering the location of the Lenghu observatory and timing of the pilot observing program from May to June, we selected Boötes III and Draco---two satellite galaxies with large angular half-light radii of 21\farcm14\xspace and 8\farcm14\xspace respectively \citep{Drlica-Wagner2022}---as test targets to fully utilize the wide field of view of WFST. 

\begin{table*}[ht!]
  \begin{center}
  \caption{Key parameters of the WFST observations for Boötes III and Draco}
  \label{tab:obs}
  \begin{tabular}{cllcllll}
  \hline\hline\noalign{\smallskip}
  Target & \ra  & \ensuremath{\mathrm{Decl.}}  & Filter & Exp. time  & FWHM  & Start Time & Sky background brightness \\
  & deg & deg &  & s & \asec & UTC & \mac \\
  \hline\noalign{\smallskip}
  Boötes III  & 209.300 & 26.800 & $g$ & 25$\times$120 & 1.56 & 2024-05-15 19:30:42 & 20.26\\ 
  Boötes III  & 209.300 & 26.800 & $r$ & 25$\times$120 & 1.32 & 2024-05-13 19:31:37 & 19.94\\
  Draco  & 260.052 & 57.915 & $g$ & 20$\times$180 & 1.48 & 2024-06-06 18:38:04 & 19.71\\
  Draco  & 260.052 & 57.915 & $r$ & 20$\times$180 & 1.55 & 2024-06-07 18:33:26 & 20.20\\
  \noalign{\smallskip}\hline
  \end{tabular}
  \end{center}
  \end{table*}

For each set of images with a total cumulative exposure time of $\roughly 1$\,hour, we measure the full width at half maximum (FWHM) and ellipticity of the stellar point spread functions (PSFs) using \sextractor and \PSFEx. Additionally, we calculate the sky background brightness, and the results are summarized in \tabref{obs}. In general, the Boötes III images have a median FWHM of 1\farcs56\xspace and a median sky background brightness of 19.94\mac in the $g$ band, while the $r$-band images have a median FWHM of 1\farcs32\xspace and a median sky background brightness of 20.26\mac. For the Draco field, the median FWHM is 1\farcs48\xspace in the $g$ band, and 1\farcs55\xspace in the $r$ band, with corresponding median sky background brightness of 19.71\mac and 20.20\mac. The ellipticity of the stellar PSFs in all four sets of images is generally below 0.1.

\section{Data reduction} 
\label{sec:reduction}
     
\subsection{Detrend Process} 
\label{sec:ccd calibration}
Ensuring a consistent and uniform detector response across the $\roughly 6.5 \deg^{2}$ field of view of WFST is essential for obtaining deep stacked images \citep{Waters2020}. In brief, this calibration is performed through the following steps:
\begin{enumerate}
  \item For each channel in bias, flat and science images, an overscan value is substracted, calculated as the median of the pixels in the overscan region. Then, the 144 channel images are concatenated into 9 CCD images for single epoch.
  \item We calculate the mean values of each pixel across 20 bias images to generate a master bias frame. This master bias is then used to correct the pixel-to-pixel variations in the science images caused by the bias structure. 
  \item Similarly, we normalize every flat images within the linear response regime by dividing them by the median value of pixels in each central CCD image. The flat images used for this calibration are twilight flat taken on the same day as science images. All normalized flat images are combined to create a master flat frame, which is then applied to our science images to correct for  photon response variations pixel by pixel across the focal plane.
  \item Finally, we apply the ``star-flat" method to correct for spatial variations in the photometric zero point across the focal plane for every science images. This method fits a position-dependent zero point using reference magnitudes and instrumental magnitudes. The correction is performed after the initial photometric calibration, with a detailed description provided in \secref{photometry}.
\end{enumerate}

\subsection{Astrometric Calibration}
\label{sec:astrometry}
We calculate our initial astrometric solution by using \textit{astrometry.net}  \footnote{\url{https://github.com/dstndstn/astrometry.net}} to insert standard WCS keywords to the header of each CCD image. Then, we offer a better astrometric solution through the following steps:
\begin{enumerate}
  \item We extract sources from individual CCD images using \SExtractor \citep{Bertin1996}, and then \PSFEx \citep{Bertin2013} is applied to the catalog to obtain a spatially variable PSF model. Finally, we obtain the position of every object based on PSF photometry by \SExtractor.
  \item Considering that the stellar positions in our catalogs are better measured using PSF model, and that the renormalised unit weight error (\code{ruwe}) provided by Gaia DR3 is expected to be close to 1.0 for sources that are well-described by a single-star model, objects with $\code{ruwe}<1.4$, $\code{pmra}/\code{pmra\_err}<5$ and $\code{pmdec}/\code{pmdec\_err}<5$ in \Gaia DR3 are loaded as the stellar reference catalog. Then, a position correction is applied to get the positions at the WFST epochs using \Gaia proper motion and parallax.
  \item Finally, The measured catalog and the reference catalog are read by SCAMP \citep{Bertin2006}. Then a fourth-order polynomial fit are applied to obtain the final astrometric solution for each CCD image.
\end{enumerate}

As a result, the standard deviations of the residuals between the calibrated stellar positions and the reference stellar positions for the four sets of images (Boötes III$-g$, Boötes III$-r$, Draco$-g$, and Draco$-r$) are 30\mas, 20\mas, 17\mas, and 15\mas respectively.

\subsection{Photometric Calibration and Image Stacking}
\label{sec:photometry}

Before stacking images, the ``star-flat" method mentioned in \secref{ccd calibration} is applied to each single-epoch CCD images to correct for non-uniform photo response caused by large-scale illumination variations in the twilight flat. For each channel, the background is modeled and subtracted using a fifth-order polynomial. Sources with PSF magnitude information are then extracted from each CCD using \sextractor and \PSFEx. We match our catalog with the PS1 DR2 catalog using a tolerance of 1\asec radius, and select stars with magnitudes between 16\magn and 18.5\magn. Given the similarity between WFST's filters \citep{Lei2023,Liu2023} and those of PS1 \citep{Chambers2018}, we calculate the magnitude residuals across the CCD, and fit a fifth-order polynomial to derive a two-dimensional zero-point correction. Finally, this zero-point correction is applied to the images without background substraction.

The image stacking process is performed using \SWarp \citep{Bertin2010} after the background subtraction. Source extraction and photometry are then conducted with \SExtractor and \PSFEx, following the same procedure as in step1 of \secref{astrometry}. Finally, the calibrated catalogs from the stacked images are obtained through differential photometry, using the PS1 MeanObject catalog as a reference.

\section{Member stars selection}
\label{sec:selection}
The selection of member stars is crucial for the subsequent analysis of Boötes III's structure and the PM measurements of both satellite galaxies.  Since dwarf galaxies contain fewer stars than typical galaxies, contamination from foreground stars and background galaxies has a more significant impact on our results.

\subsection{Completeness and Purity}
\label{sec:completeness}
Before selecting member stars, we estimate the completeness and purity (the fraction of real sources among all detected sources) of our catalogs to assess the photometric depth and make a reliable magnitude cut. Typically, in large survey projects, photometric depth is assessed by injecting artificial sources into real images and re-detecting them across a large number of exposures \citep{Metcalfe2013}. However, it is not the aim of this paper to systematically evaluate the photometric depth. Instead, we adopt a simpler but effective approach by comparing our catalogs with deeper reference catalogs to estimate completeness and purity. \cite{segal2007} carried out $g$-, $r$-, and $i$-band observations of a $5 \deg^{2}$  region around the Draco galaxies using the CFHT MegaCam, with the seeing ranging from 0\farcs461\xspace to 1\farcs121\xspace. Their catalogs achieves a completeness of over 50\% for sources with $i < 24.5\magn$. We use these catalogs to achieve our goals for catalogs in the Draco field. For the Boötes III catalogs, as no deeper catalogs with completeness analysis exist beyond our limiting magnitude, we estimate its completeness using the relationship between completeness and the turnover point in source counts \citep{Metcalfe2013}.

\begin{figure}[ht!]
  \center
  \includegraphics[width=\columnwidth]{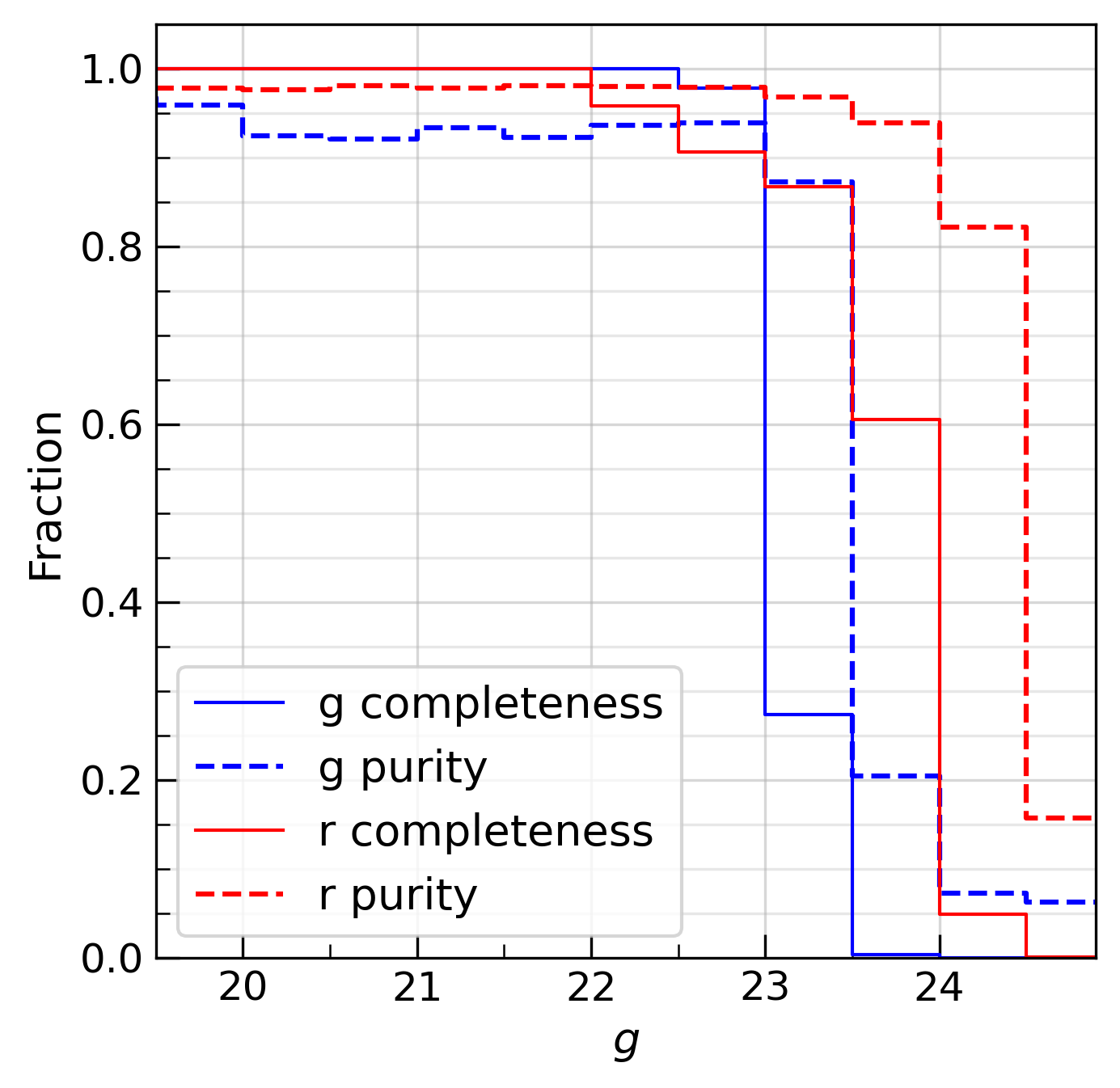}
  \caption{The completeness and purity of $g-$ and $r$-band sources in the Draco field across different magnitudes.}
  \label{fig:comp}
  \end{figure}

We exclude spurious sources with $\code{class\_i}=0$ from the \cite{segal2007} catalog and match the observation regions, totaling $3.9 \deg^{2}$. \figref{comp} shows the completeness and purity of the WFST $g$-band and $r$-band catalogs. The $g$-band catalog achieves a completeness over 90\% for stars brighter than 23\magn, while the $r$-band catalog achieves a completeness over 60\% for stars brighter than 24\magn. Notably, since our sample selection is color-based, the final completeness is limited by the catalog with the lower completeness. In addition, cross-matching the $g$-band and $r$-band catalogs enhances overall catalog purity, as most contaminants, upon visual inspection, are located near bright stars. Finally, for the Draco field,  we cross-match $g$-band and $r$-band catalogs with a radius of 1\asec, and apply a magnitude cut of $g < 23\magn$ to the catalog. 

Figure\,14 of \cite{Metcalfe2013} shows a relation between the predicted 5-sigma limiting magnitude (equal measured 50\% completeness limiting magnitude) and the measured number counts turnover. They found a systematic offset ($\roughly 0.6\magn$) between these two magnitudes. As the measured number counts turnover represents the balance point between the increasing number of fainter sources and the decreasing detection efficiency at the faint end, this discrepancy came partly from the lower detection efficiency for galaxies and partly because, the count peak occurs at a magnitude somewhat brighter than the 50\% limiting magnitude.

\begin{figure}
\center
\includegraphics[width=\columnwidth]{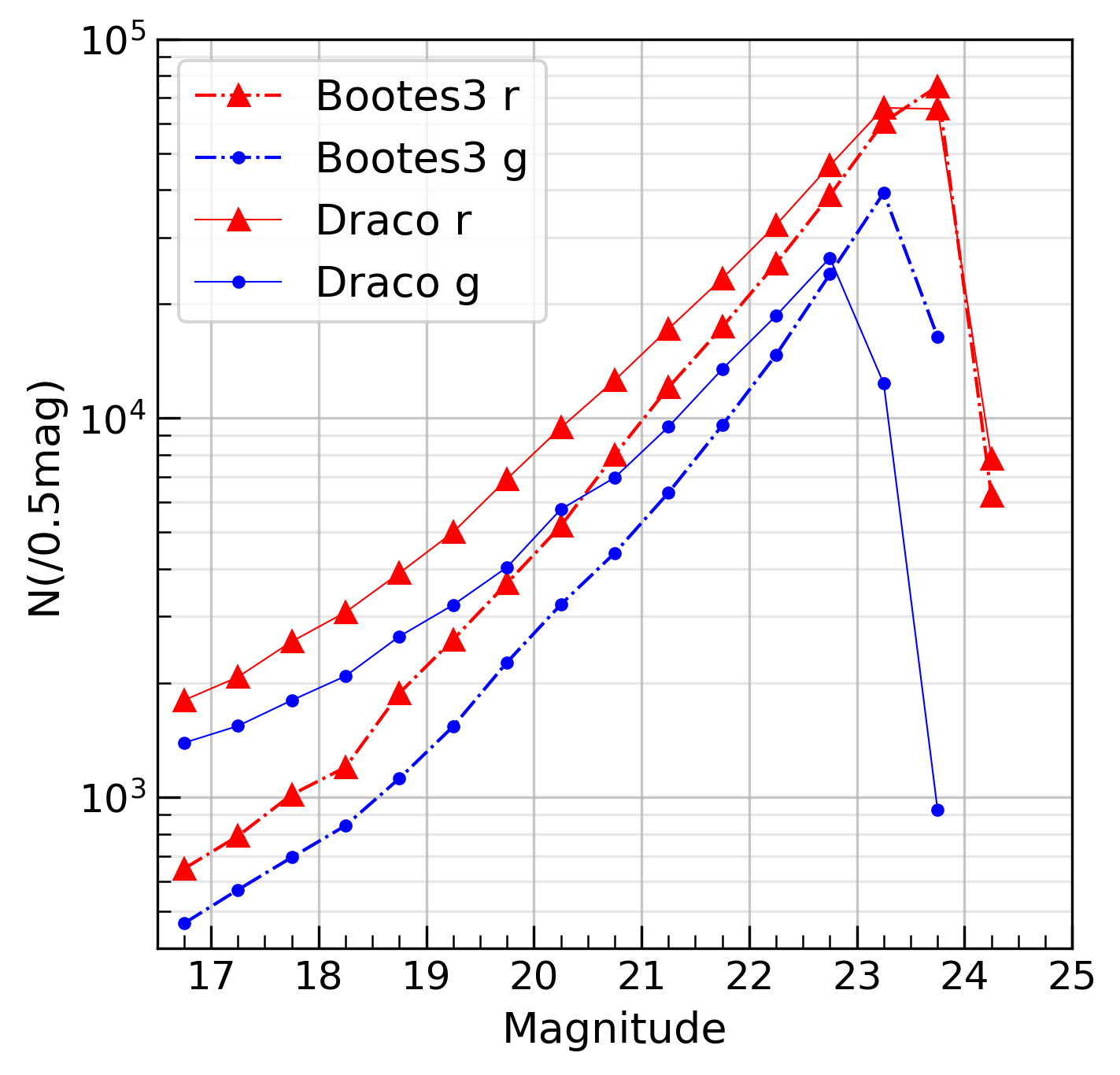}
\caption{The number of sources in 0.5\magn bin for Draco and Boötes III fields. The solid line represents the Draco field, while the dash-dotted line represents the Boötes III field.}
\label{fig:scounts}
\end{figure}

\figref{scounts} shows the number of detected sources as a function of magnitude for our catalogs. In the Draco field, the turnover magnitude in the source counts distribution ($\roughly$22.5--23\magn and 23--23.5\magn in $g$-band and $r$-band) lies at a brighter magnitude than that corresponding to 50\% completeness. This result is consistent with the findings of \cite{Metcalfe2013}. Given that the same data processing pipeline was applied to all images, and the FWHM of PSF is similar between the Draco field and Boötes III field in the corresponding bands, we conclude that, in the Boötes III field, the $g$-band catalog achieves a completeness over 50\% for stars brighter than 23.5\magn, while the $r$-band catalog achieves a completeness over 50\% for stars brighter than 24\magn. Based on this estimation, we cross-match $g$-band and $r$-band catalogs with a radius of 1\asec and exclude sources with $g > 23.5\magn$ from the final cross-matched catalog. 

\subsection{Star and Galaxy Separation}
   
\begin{figure*}
\center
\includegraphics[scale=2, width=\textwidth]{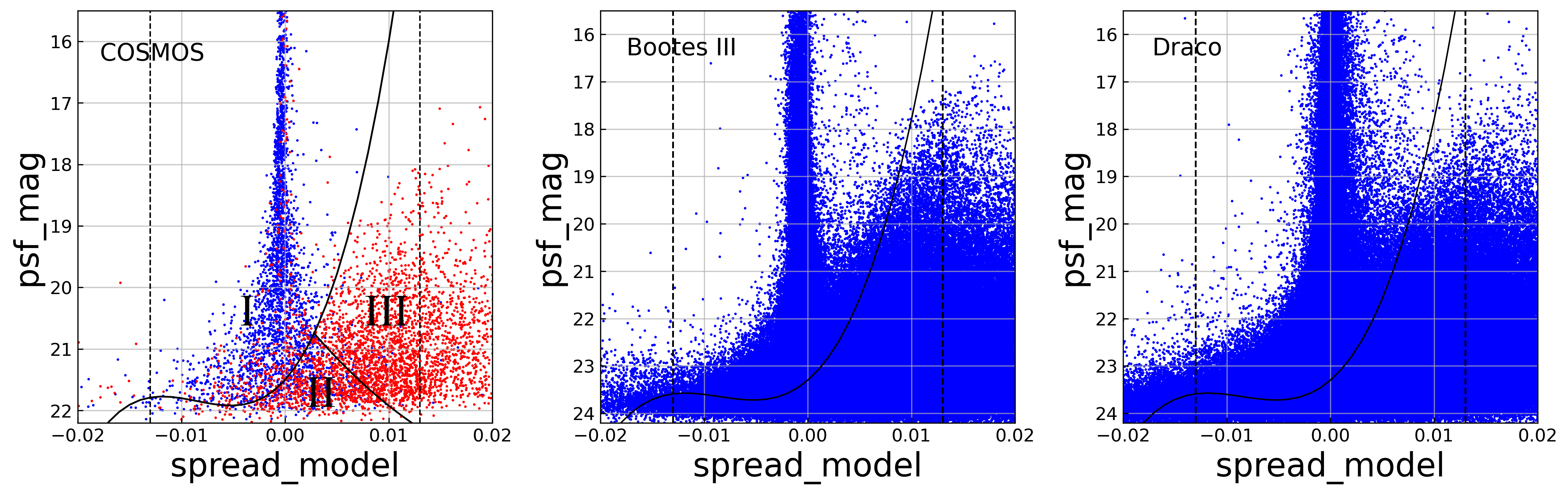}
\caption{The \magpsf axis shows psf magnitude for all detected sources in the $r$-band. The \spreadmodel parameter is shown on the x-axis, with the vertical dashed lines marking the selection boundaries. The left panel corresponds to the COSMOS field, where the solid lines divide the plot into three regions. The fractions of sources classified as stars in regions I, II, and III are 86.3\%, 32.8\%, and 5.5\%, respectively.} The middle panel represents the Boötes III field, and the right panel represents the Draco field.  In our selection criteria, we include sources that are located above the solid line and within the two dashed lines.
\label{fig:star}
\end{figure*}

As stars are the fundamental building blocks of Milky Way satellite galaxies, star-galaxy (S/G) separation plays an important role in identifying and mapping the MW satellite galaxies. For example, \cite{Homma2024} applied a new S/G separation method to a large amount of HSC data, and identified a new Milky Way satellite galaxy, Virgo 3. 

In general, S/G separation for astrophysical sources includes morphology-based methods \citep{slater2020, henrion2011,Farrow2014}, which classify sources as stars by evaluating their similarity to the PSF model. Additionally, methods based on multi-band photometry or spectral energy distributions \citep{fadely2012, cook2024} have been developed, utilizing machine learning algorithm to estimate the probability of a source being a star. More recently, Convolutional Neural Networks (CNNs)-based approaches have been applied to images to directly distinguish stars from galaxies \citep{stoppa2023}. With a sufficiently large WFST dataset in the future, these three kinds of methods can all be tested and compared. However, given the limited dataset from the pilot observing program, we adopt a method similar to that of \cite{koposov2015} to achieve our goals. Specifically, we measured the \spreadmodel \footnote{\url{https://sextractor.readthedocs.io/en/latest/Model.html}} parameter for each source using \SExtractor. To minimize variations in instrumental effects and observational conditions on \spreadmodel, we apply the same processing pipeline on the COSMOS field observed by WFST on a adjacent night. The S/G separation criterion is established based on the distribution of sources in the $\var{mag\_psf}_\texttt{r}-\spreadmodel$ diagram, as shown in \figref{star}. As estimate from the left panel, the fraction of sources classified as stars is \roughly 75\% at the limiting magnitude of the Boötes III catalog, and \roughly 85\% at the limiting magnitude of the Draco catalog. For the PM sample discussed in \secref{faint selection}, this fraction increases to over 90\% in both fields. It should be noted that for faint sources, some stars may be excluded. However, the overall fraction of stars in this magnitude range increases, which better aligns with our goal of studying the stellar distribution in these two galaxies.

\subsection{Decontamination with \Gaia}
\Gaia Data Release 3 (\Gaia DR3), which has been released on 13 June 2022, provides classifications (stars, galaxies, quasars) for $\roughly 1.6$ billion sources and stellar parameters for about 470 million stars \citep{Gaia2023}. Additionally, \Gaia offers positions and PMs for 1.468 billion sources \citep{Lindegren2021}, enabling an independent method for excluding foreground stars. Furthermore, background galaxies can be effectively filtered out due to \Gaia's space-based observations, which have an effective angular resolution of $\roughly$0\farcs4\xspace \citep{Gaia2018}. 
   
We cross-match our two catalogs with \Gaia DR3 using a 1\asec radius. Sources with $\code{ruwe}>1.4$ are removed from our catalogs to filter out galaxies. To eliminate contamination from foreground stars, we select sources with $\code{parallax}/\code{parallax\_error}>3$ to ensure reliable parallax measurements. Finally, we exclude sources with $\code{parallax}>0.2$, thereby foreground stars within 5 kpc from the Sun are removed.

\subsection{Color-Magnitude Diagram Selection}
\subsubsection{Boötes III}
As mentioned in \secref{intro}, MW dwarf satellite galaxies are predominantly composed of old stellar populations, allowing their member stars to be identified in a color-magnitude diagram (CMD). Before applying isochrone filtering in the CMD, we deredden all stars using the SFD98 dust map \citep{Schlegel1998, Schlafly2011}. For Boötes III, \cite{Carlin2009} estimated a metallicity of $\roughly -2.0$ based on low-resolution spectroscopy. Thus we adopt an isochrone with a metallicity of $-2.0$ and an age of 13\Gyr, shifting it to a distance of 46.5\kpc \citep{Carlin2018} to select member stars. 

\begin{figure*}
\centering
\includegraphics[width=1\textwidth]{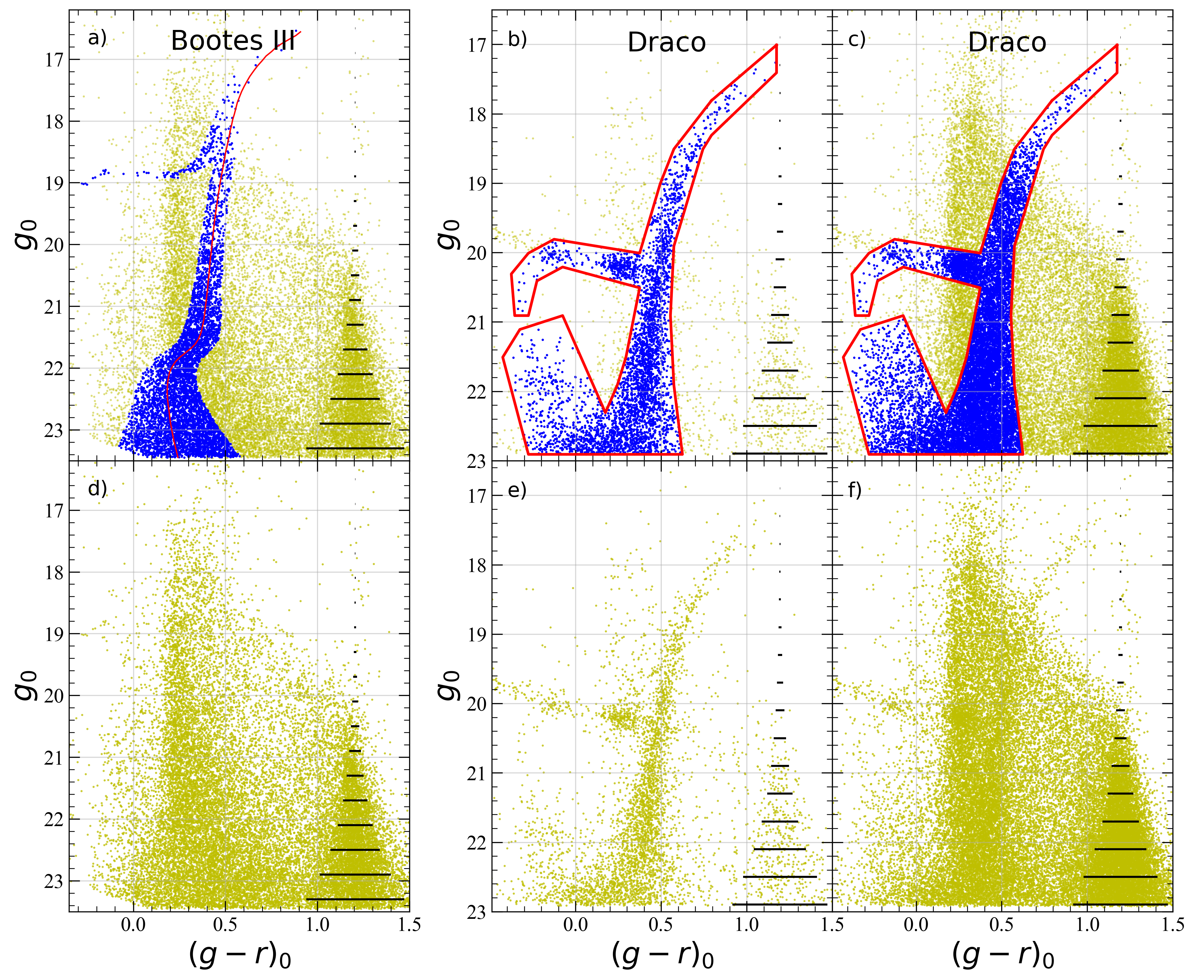}
\caption{Panel~a) shows the color-magnitude diagram (CMD) of Boötes III. The red solid line represents an isochrone with an age of 13\Gyr and a metallicity of $-2.0$. Blue dots represent selected stars from our match-filter method. Panel~b) shows the stars within a 16\amin radius of Draco, while Panel~c) represents all detected stars in the Draco field. The red lines in panels b) and c) indicate the selection boundaries used to identify candidate member stars. The photometric uncertainties are illustrated as black solid lines on the right side of each panel. The bottom three panels shows the corresponding stars without color separation in each fields.}
\label{fig:match filter}
\end{figure*}

Given the photometric uncertainties in both color and magnitude, we extend the filtering region along the color axis based on the color uncertainty at each $g$-band. Additionally, a 0.05\magn is applied to the color axis to account for intrinsic color variation of Boötes III stars and uncertainties in dereddening corrections. Similarly, for horizontal branch stars, a 0.05\magn is added along the magnitude axis to account for the effects of the physical extent of Boötes III and the uncertainties in its distance measurement. The final selection diagram is shown in panel~a) of \figref{match filter}.
   
\subsubsection{Draco}
Draco has a smaller angular half-light radius (8\farcm14\xspace) compared to Boötes III (21\farcm14\xspace), and its stellar population is clearly distinguishable in the CMD. Thus, we select all the sources within a 16\amin radius centered on Draco and use the distribution of these sources in the CMD to filter sources across the entire field of view. The results are shown in \figref{match filter}. Similar to the results of \cite{gall2007}, our selected stars include RGB, AGB, and horizontal branch stars. However, due to the depth of our observations, the main-sequence stars are not included.

\subsection{Proper Motion Selection}

\subsubsection{Bright Star selection}
\begin{figure*}[ht]
\center
\includegraphics[width=\textwidth]{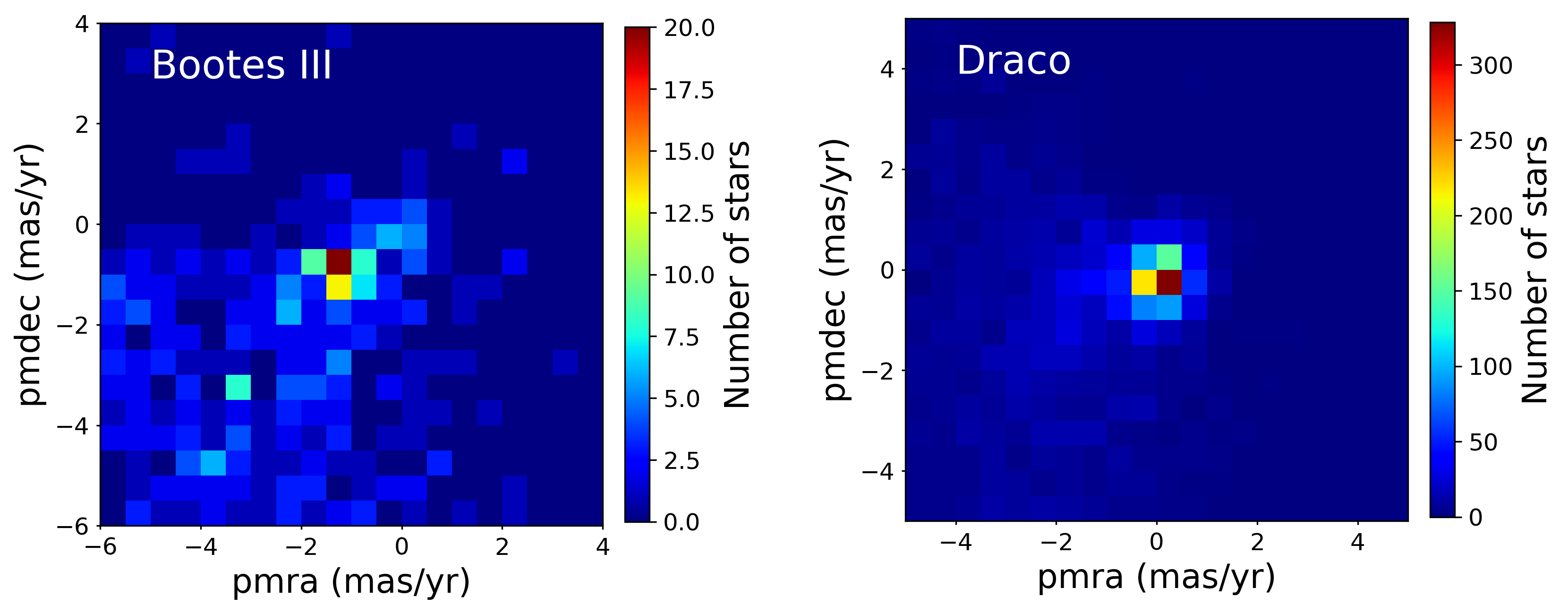}
\caption{Spatial distributions of stellar PMs for bright stars in Boötes III (left) and Draco (right) field. The bin size is $0.5\masy\times0.5\masy$}
\label{fig:pm}
\end{figure*}

With a 34-month time baseline, \Gaia DR3 improves precision of PM measurements by a factor of \roughly2 compared to \Gaia DR2 (Figure\,2 in \cite{Gaia2023}). This improvement provides more precise and independent dimensions (\pmra and \pmdec) for filtering out foreground stars and characterizing the kinematics of the satellite galaxy. First, we select stars with $g$-band magnitudes brighter than 21\magn in our catalogs. Then, we exclude stars lacking \Gaia PMs ($\roughly 3.5\%$ for the Boötes III catalog and $\roughly 1.7\%$ for the Draco catalog). The remaining stars are binned into 0.5\masy intervals in both \pmra and \pmdec dimensions, as shown in \figref{pm}. 
 
For Boötes III stars, The central region at $\pmra=-1.25$\masy and $\pmdec=-0.75$\masy exhibits a density excess more than four times that of the surrounding areas. For Draco stars, this excess is even more pronounced, with the central region at $\pmra=0.25$\masy and $\pmdec=-0.25$\masy reaching over nine times the surrounding density. Using the PMs of stars in these bins, we apply the trim-mean algorithm, obtaining mean uncertainties of 0.226\masy for \pmra and 0.176\masy for \pmdec in the Boötes III field and 0.308\masy for \pmra and 0.353\masy for \pmdec in the Draco field. Considering the uncertainties in \Gaia PMs and the intrinsic PM dispersion of member stars in satellite galaxies, we smooth the data using a Gaussian kernel with a sigma of approximately three times the mean PM uncertainty. Finally, we calculate the mean stellar count within the central 20\% of the distribution, and exclude stars in bins where the smoothed stellar counts fall below three times of this mean.
  
\subsubsection{Faint Star selection}
\label{sec:faint selection}
\begin{figure*}[th!]
  \center
  \includegraphics[width=\textwidth]{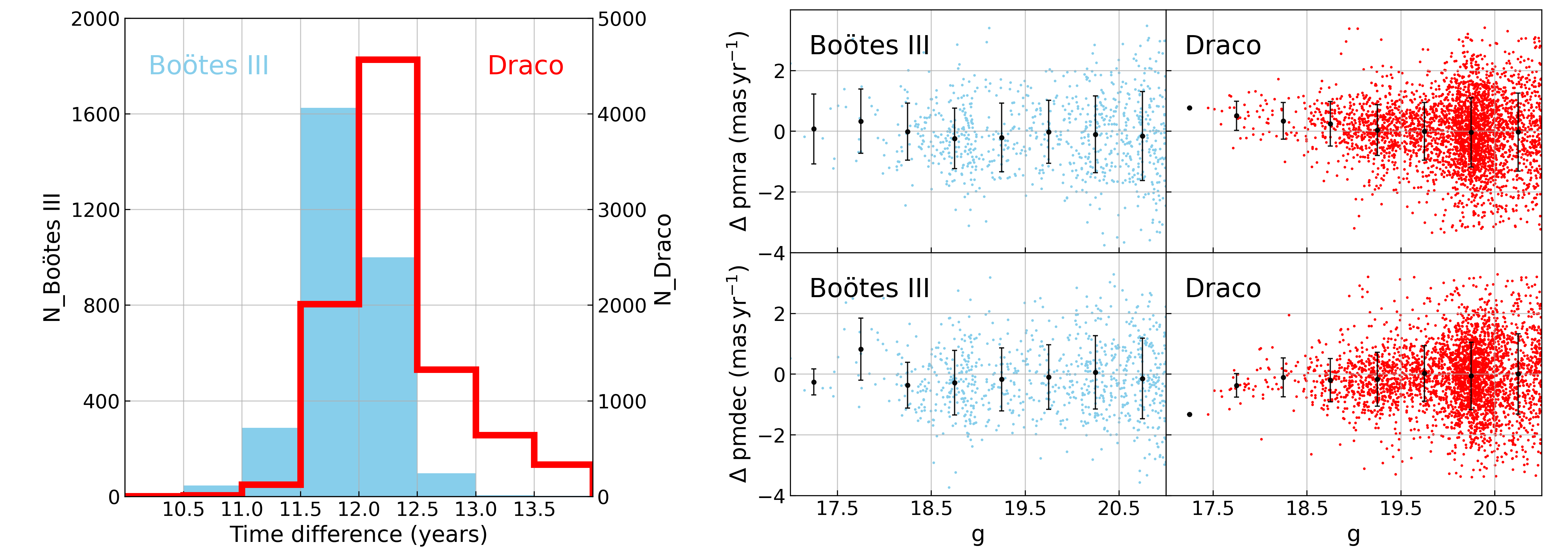}
  \caption{The left panel shows the distribution of the time baseline for selected stars that have our proper motion measurements. The four panels on the right present the differences between our PMs and those from \Gaia. Sky-blue dots represent the candiate member stars of Boötes III, while red dots correspond to the candiate member stars of Draco. Black dots indicate the mean values within a 0.5\magn intervals, and error bars represent the standard deviation.}
  \label{fig:time}
\end{figure*}

For faint stars with $g>21\magn$, we measure PMs by combining our catalogs with PS1 DR2 over a 12-year time baseline. The PS1 AstrometryCorrection table \citep{lubow2021,white2022}, recalibrated in 2022 based on \Gaia EDR3, achieves a median residual of 9\mas between PS1 and \Gaia positions. In their calibration, each \Gaia stellar position was shifted to the epochs of PS1 stellar position by applying stellar parallax and PM. An astrometric reference frame was then constructed with corrections for systematic uncertainties introduced by small-scale geometric distortions, stellar color terms, differential chromatic refraction in \textbf{declination}. As discussed in \secref{astrometry}, our astrometry solution is based on the \Gaia DR3, which shares identical astrometric data with \Gaia EDR3. Therefore, stellar positions in both calatogs are defined within the same reference coordinate system. Based on this, We then calculate the PMs of the selected stars by the following procedure:
\begin{enumerate}
  \item  We cross-match the two catalogs within a radius of 1\asec and derive \pmra and \pmdec by 
    \begin{equation}
    \scriptsize
    \begin{split}
    \pmra &= \frac{(\mra_{\rm WFST}-\mra_{\rm PS1})\times \cos ({\rm \mdec}_{\rm WFST})}{{\rm epoch}_{\rm WFST}-{\rm epoch}_{\rm PS1}}+C_1 \\
    \pmdec &= \frac{(\mdec_{\rm WFST}-\mdec_{\rm PS1})}{{\rm epoch}_{\rm WFST}-{\rm epoch}_{\rm PS1}}+C_2
    \end{split}
    \label{eqn:pm}
    \end{equation} 
   
    The time differences between observations are shown in \figref{time}.
  \item $C_1$ and $C_2$, initially set to zero, are constants for systematic corrections of PM. We use stars with $g <21\magn$ in each catalog to derive these two constants. 
  \item We calculate the residuals of the \pmra and \pmdec derived from the above steps and \Gaia DR3. Given the limited WFST data, obtaining a reliable correction for systematic uncertainties in astrometric calibration, as done for PS1, remains challenging. These systematic uncertainties inevitably propagated into the PM calculations. To mitigate these systematic uncertainties, we apply the mean residuals as $C_1$ and $C_2$ to derive the final PMs.
\end{enumerate}

\begin{figure*}
  \center
  \includegraphics[width=\textwidth]{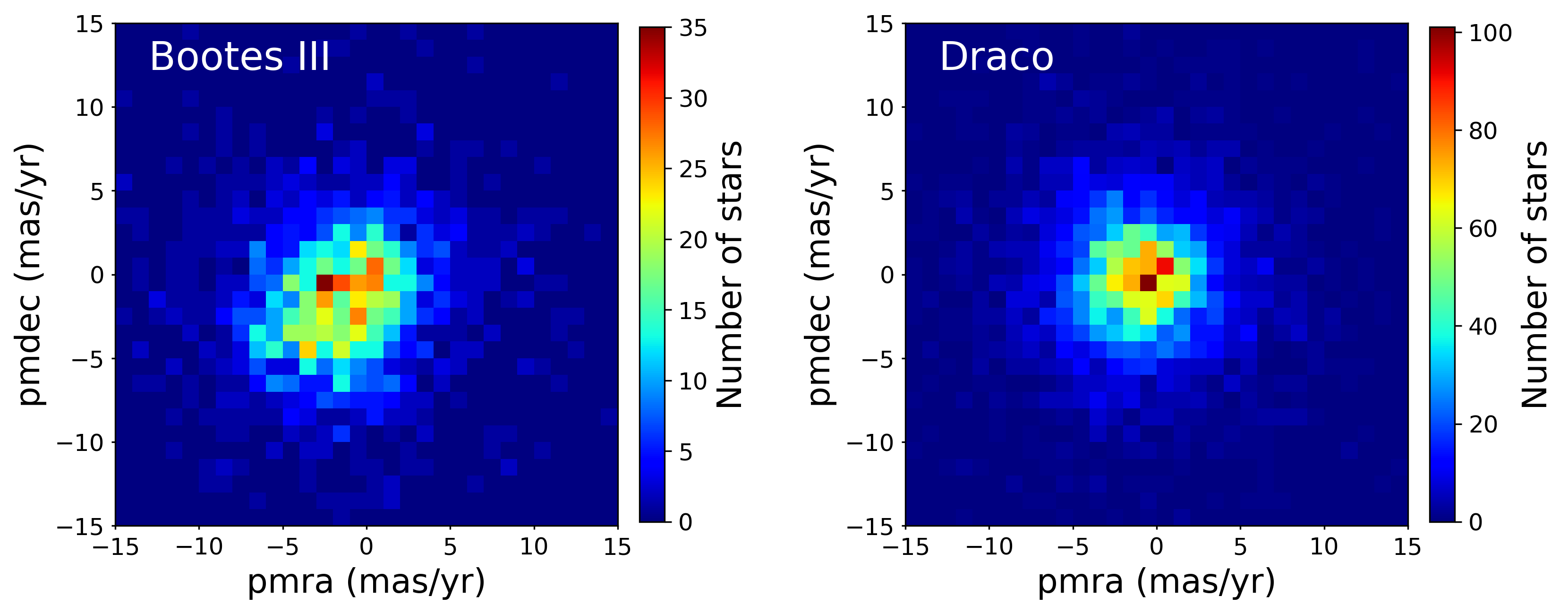}
  \caption{Spatial distributions of stellar PMs for faint stars in Boötes III (left) and Draco (right) field. The bin size is $1\masy\times1\masy$}
  \label{fig:pm_wfst}
  \end{figure*}

We obtain $C_1=-0.38$\masy and $C_2=0.28$\masy in the Boötes III field and $C_1=-0.12$\masy and $C_2=0.32$\masy in the Draco field. As shown in \figref{time}, the standard deviation of the residuals is 1.21\masy for \pmra and 1.16\masy for \pmdec in the Boötes III field and 1.09\masy for \pmra and 1.08\masy for \pmdec in the Draco field. The peak stellar density of faint stars occurs in the same bin as the peak stellar density of bright stars. Similarly, in the PM density maps for all stars, the high-density regions are concentrated near these bins. To further reduce the contamination of faint foreground stars, stars with $21 <g < 22.5\magn$ were binned into 1\masy intervals in \pmra and \pmdec over the range of $-15$\masy to 15\masy, \textbf{as shown in \figref{pm_wfst}.} A Gaussian kernel with a size of 1 pixel is applied to smooth the density map, which consists of $30\times30$ pixels. Pixels with stellar density below a certain threshold are excluded, and the remaining stars are identified as the candidate member stars.

\section{Results and Discussion}
\label{sec:results}

\subsection{Proper Motion}

\subsubsection{Uncertainty}
We calculate the PM uncertainty ($\sigma$) for each coordinate using \eqnref{pmerr}:
\begin{equation}
\scriptsize
{\sigma}^2 = \frac{{\sigma}^2_{{\rm P}_{\rm WFST}}+{\sigma}^2_{{\rm P}_{\rm PS1}}}{t^2}+{\sigma}^2_C+\frac{({\rm P}_{\rm WFST}-{\rm P}_{\rm PS1})^2}{t^4}\times{\sigma}^2_{\rm t}+{\sigma}^2_{\rm ex}
\label{eqn:pmerr}
\end{equation} 
where ${\sigma}_{{\rm P}_{\rm WFST}}$ represents the positional measurement uncertainty calculated as \errpsf using \sextractor, and ${\sigma}_{{\rm P}_{\rm PS1}}$ corresponds to the positional measurement uncertainty from PS1 AstrometryCorrection table. ${\sigma}_C$ accounts for the systematic uncertainty in \Gaia PM measurements \citep[Table~7 of ][]{Lindegren2021}. The third term in the equation is negligible, as ${\sigma}_{\rm t}$ is significantly smaller than the others. The additional uncertainty term, ${\sigma}_{\rm ex}$, is derived in 0.5\magn bins by computing the difference between the standard deviation of the residuals and the sum of the first two terms. For ${\sigma}_{\rm ex}$ of fainter stars, we adopt the mean ${\sigma}_{\rm ex}$ estimated from bright stars. The additional uncertainties arising from small-scale distortions and differential chromatic refraction are independent of magnitude \citep{Lin2020}, while other effects (e.g., charge transfer inefficiency, CTE) may exhibit magnitude dependence. This suggests that our uncertainty estimates for faint stars could be underestimated.

\begin{figure}
\center
\includegraphics[width=\columnwidth]{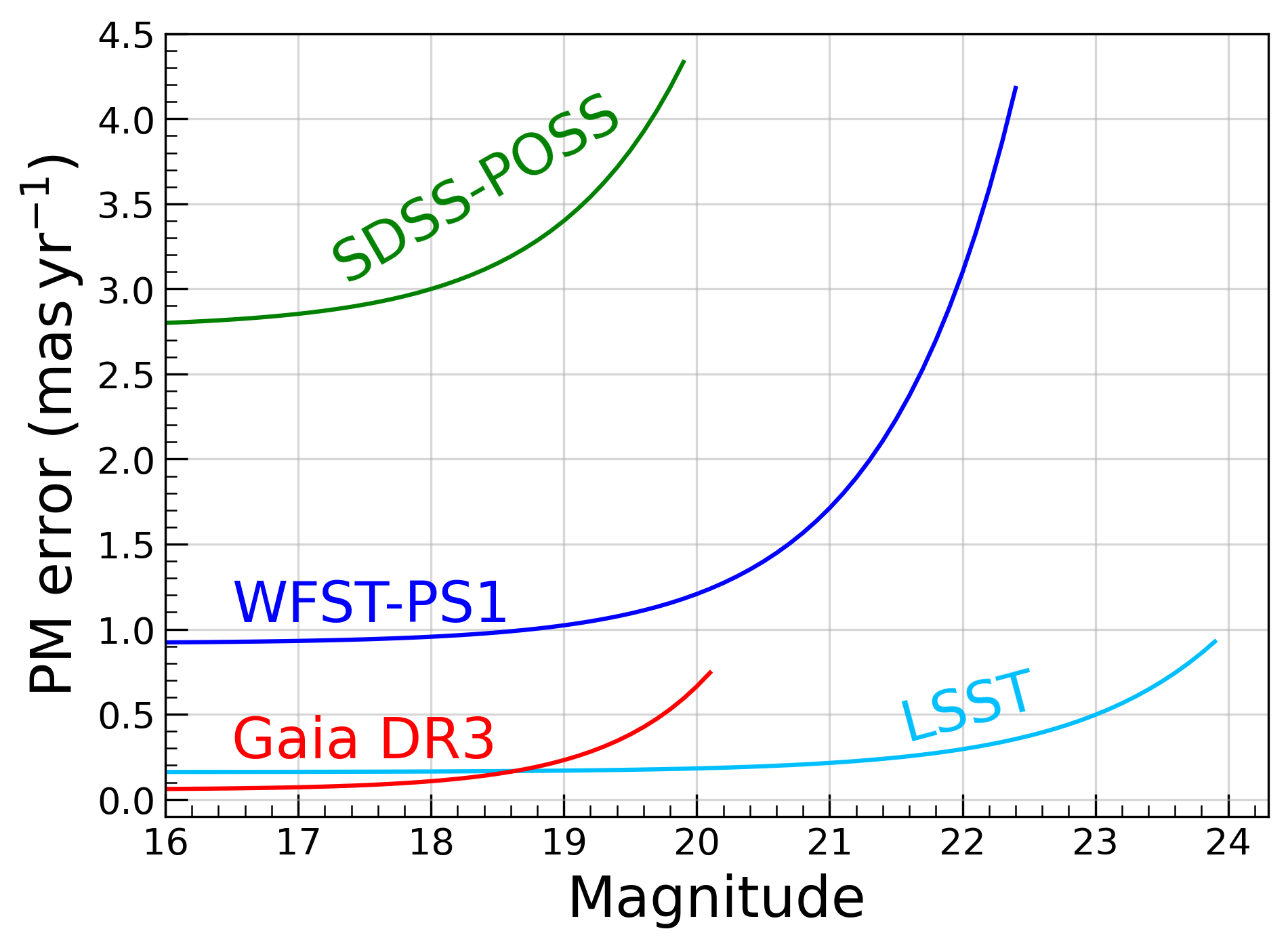}
\caption{Illustration of the variation in proper motion uncertainty per coordinate as a function of magnitude. The green solid line represents the SDSS-POSS $r$-band data \citep{Munn2004}, while the deep sky-blue solid line corresponds to the LSST $r$-band test data \citep{Rich2018}. The red line denotes the \Gaia DR3 $G$-band data in the Boötes III and Draco regions, and the blue line represents the results obtained by combining WFST data with PS1 DR2 in the $r$ band}
\label{fig:pm_err}
\end{figure}

We present the results in \figref{pm_err}. The mean PM uncertainty, derived by combining our catalog with PS1 DR2, is approximately 1.8\masy for stars with $r = 21\magn$, and increases to 3.0\masy for stars with $r = 22\magn$. Compared to other surveys, WFST-PS1 PM measurements provide a unique dataset for stars with $r>21\magn$.

\subsubsection{Boötes III}
\begin{figure*}
\center
\includegraphics[width=\textwidth]{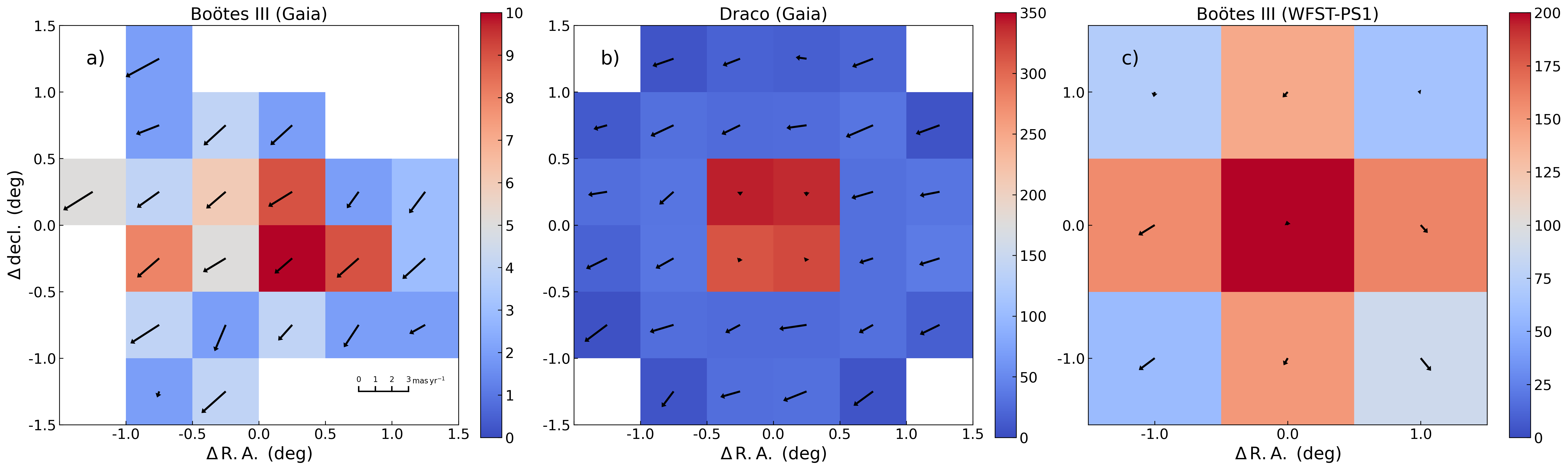}
\caption{The left and middle panels show binned density maps for candidate member stars with $g$-band magnitudes brighter than 21\magn, where the color of each bin represents the number of stars, and the black arrow indicate the mean \Gaia PM for stars within each bin. The right panel displays the binned density map for candidate member stars fainter than 21\magn in the $g$ band, with arrow representing residual PM, calculated as the difference between the mean PM of stars in each bin and the PM of Boötes III. The scale bar in the left panel indicates the reference length for the proper motion vectors shown across all three panels.}
\label{fig:gaia pm}
\end{figure*}

For the candiate member stars of Boötes III, 103 bright stars with $g < 21\magn$ have PM measurements from \Gaia. In comparison, 1,118 faint stars with $21 <g < 22.5\magn$ have independent PM measurements from our calculation. We divide stars with \Gaia PM measurements into spatial bins of $0.5\degree\times0.5\degree$ and calculate the mean PM for bins containing more than one star. The left panel of \figref{gaia pm} presents the density map of Boötes III, with PMs indicated for each bin. Bins with high star counts and similar PMs are concentrated at the center, highlighting an overdensity in the core of  Boötes III. Assuming that the uncertainty of PM for each individual star follows a normal distribution, we apply a Monte Carlo method to derive the PM of Boötes III based on bright stars. In each iteration, a set of PMs is generated based on the measured values and their uncertainties, and the mean PM is calculated. The process is repeated 10,000 times. The final PM of Boötes III is determined as the mean of the simulated means, with the measurement uncertainty given by their standard deviation. Using \Gaia PM measurements, we derive the PM of Boötes III as $\pmra=-1.26\pm0.05$\masy, $\pmdec=-1.05\pm0.04$\masy. The intrinsic dispersion of PMs, calculated as the mean standard deviation of simulated values, is 0.77\masy for \pmra and 0.63\masy for \pmdec.
   
For faint stars with our PM measurements, We divide them into spatial bins of $1\degree\times1\degree$, as shown in the right panel of \figref{gaia pm}. Then, we calculate the mean PM of the stars for each bin using a trim-mean algorithm, which removes the top and bottom 10\% of the data before calculating the mean value. The PM vectors shown in \figref{gaia pm} represent the residual PMs. They are obtained by subtracting the PM of Boötes III, as estimated from bright stars, from the mean PMs in each bin. The vector in the central bin is significantly shorter, indicating that the faint stars in this region have PMs consistent with those of the bright stars. This provides kinematic evidence for the presence of a central core in Boötes III. Finally, we apply the same method to the faint stars in the central bin (204 stars) and derive the PM for Boötes III of $\pmra=-1.31\pm0.16$\masy and $\pmdec=-1.01\pm0.16$\masy, which is consistent with the PM of Boötes III calculated from PMs of bright stars.
   
\subsubsection{Draco}
\label{sec:dracop}
Same processing is applied to the final candidate member stars of Draco. 1,313 stars have PM measurements from \Gaia, and 3,258 stars have PM measurements from our calculation. The middle panel of \figref{gaia pm} represents the density map of Draco, showing that the central four bins contain a significantly higher number of bright stars with consistent PMs. Thus, we use bright stars from these four bins to derive the PM of Draco, yielding $\pmra=-0.05\pm0.01$\masy and $\pmdec=-0.24\pm0.01$\masy. And the intrinsic PM dispersion of Draco is 0.73\masy for \pmra and 0.71\masy for \pmdec. The PM of Draco based on faint stars is also derived using stars from the same region, resulting in $\pmra=-0.57\pm0.06$\masy and $\pmdec=-0.24\pm0.06$\masy. 

A summary of the proper motion measurements for these two satellite galaxies is provided in \tabref{pms}. The PMs derived from bright stars are consistent with those from faint stars, indicating no significant PM variation across stellar luminosity in these two galaxies. This result is not unexpected, as tidal interactions with the MW primarily depend on the spatial position of stars within the satellites rather than their luminosity. However, for Draco, a discrepancy of 0.52\masy of \pmra shows in the two measurements. Possible explanations include the presence of several saturated stars within the half-light radius of Draco and the alignment of the readout direction with the \ra \textbf{direction}, which may introduce systematic shifts in \ra positions that vary with stellar brightness. Contamination from foreground stars may contribute to this discrepancy. However, it is unlikely to be the primary cause of the discrepancy, as our selection process significantly reduces the fraction of foreground stars in the final catalog. Nevertheless, a more detailed validation will require a more refined astrometric solution which should be tested with a larger dataset from the WFS program. 

\renewcommand{\arraystretch}{1.2}
\setlength{\tabcolsep}{11.5pt}
\begin{table}[h]
  \flushleft
  \caption{PMs of Boötes III and Draco}
  \label{tab:pms}
  \begin{tabular}{lcc}
  \hline\hline
     & \pmra  & \pmdec  \\
     & \masy  & \masy   \\
  \hline
  Boötes III\_bright & $-1.26\pm0.05$ & $-1.05\pm0.04$\\ 
  Boötes III\_faint  & $-1.31\pm0.16$ & $-1.01\pm0.16$\\ 
  Draco\_bright & $-0.05\pm0.01$ & $-0.24\pm0.01$\\ 
  Draco\_faint  & $-0.57\pm0.06$ & $-0.24\pm0.06$\\ 
  \hline
  \end{tabular}
\end{table}

\subsection{Density Profile}
\label{sec:profile}

\subsubsection{Boötes III}
\begin{figure*}
\center
\includegraphics[width=\textwidth]{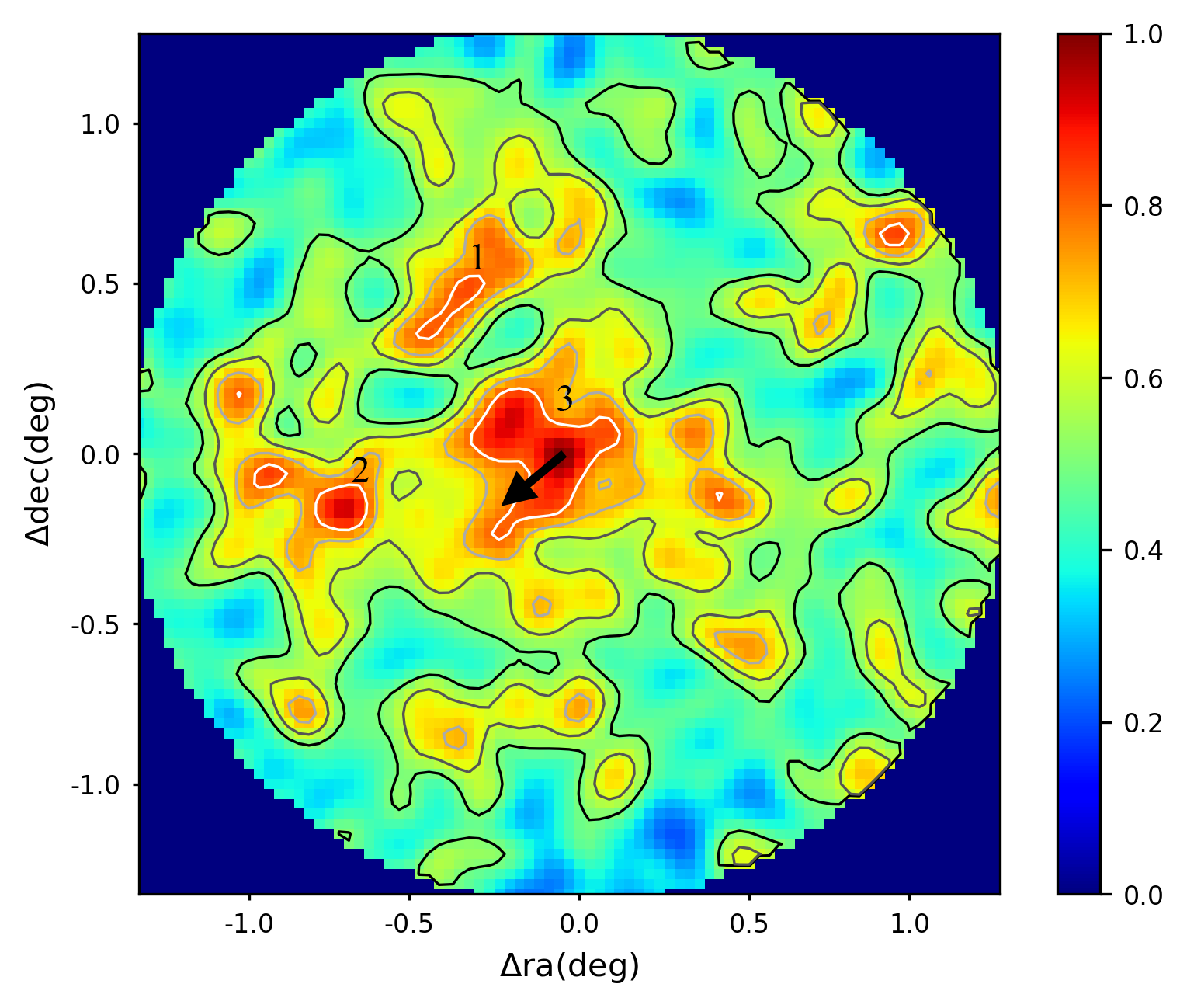}
\caption{Normalized stellar density distribution of the Boötes III galaxy. The contours correspond to density levels of 0.5, 0.6, 0.7, and 0.8. Regions 1, 2, and 3 denote three distinct high-density areas. And the black arrow at the center indicates the PM of Boötes III galaxy. }
\label{fig:density}
\end{figure*}

The candidate member stars of Boötes III within the $3\degree\times3\degree$ region are binned into $100\times100$ grids. The resulting stellar density map is then smoothed using a gaussian kernel with a sigma of 2 pixels. \figref{density} presents the final normalized density map, where three distinct regions with density levels exceeding 0.8 are identified from the contours. Region\,1 exhibits an elongated structure along with PM of Boötes III. The alignment of Regions 2 and 3 along the PM direction suggests that Region\,2 has been tidally stripped from the core of Boötes III due to the tidal force of the MW. Additionally, the elongated shape of Region\,3 indicates that the core of Boötes III is also undergoing tidal stretching. The counter with density level of 0.6, which spans a larger scale, further supports this interpretation. It suggests that Region\,2 represents the leading part of the tidal structure originating from Region\,3. We further convolve the stellar density map using gaussian kernels of varying sizes, using sigma values ranging from one to four pixels. While smaller kernels results in more fragmented substructures in Regions 2 and 3, and larger kernels produce smoother structures, the elongated structures in both regions remain visible  across all smoothing scales.

\subsubsection{Draco}
\begin{figure}
  \center
  \includegraphics[width=\columnwidth]{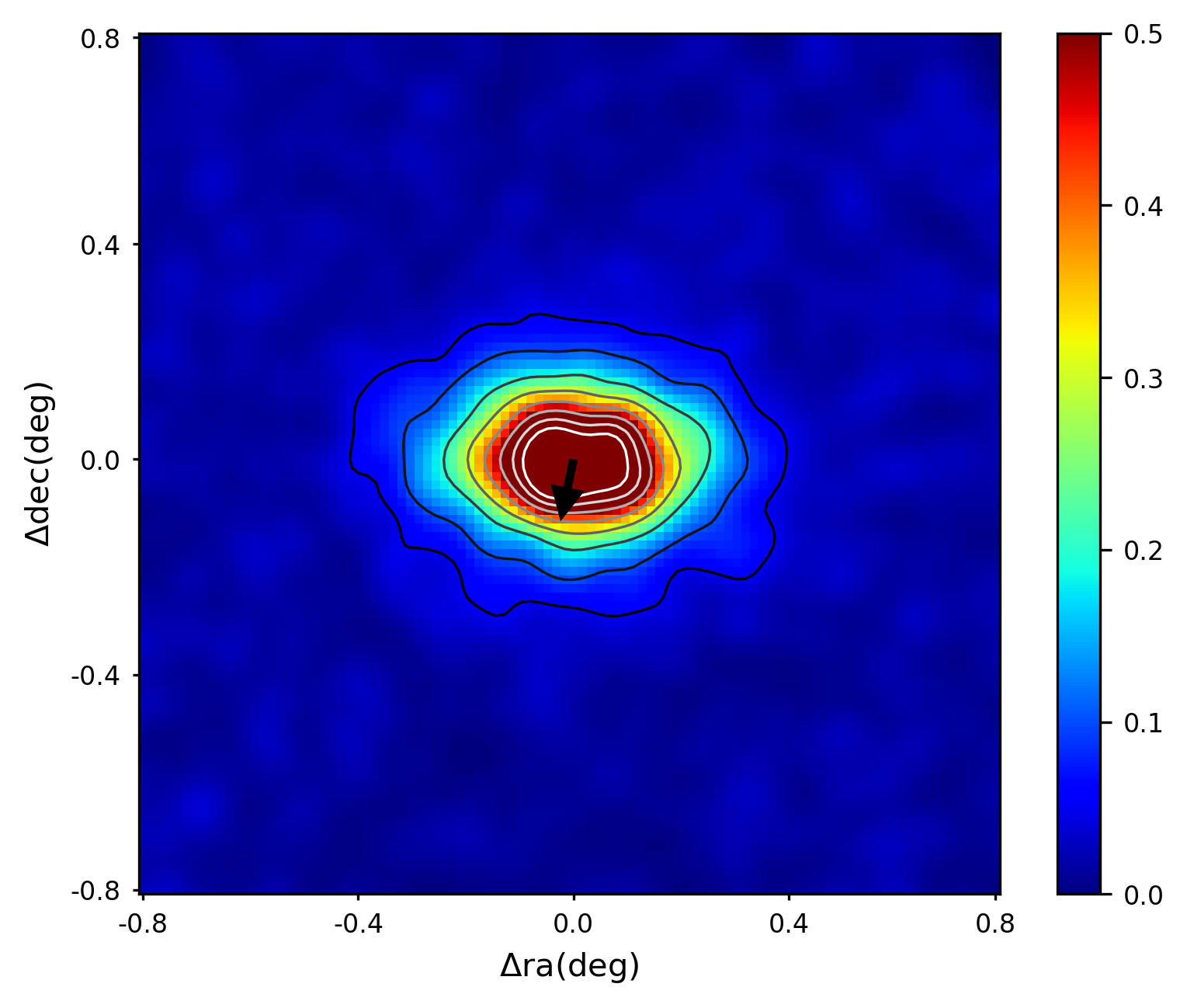}
  \caption{Normalized stellar density distribution of the Draco galaxy. The contours correspond to density levels of 0.05, 0.1, 0.2, 0.3, 0.4, 0.5, 0.6, 0.7. The black arrow at the center indicates the PM of Draco galaxy. }
  \label{fig:ddensity}
  \end{figure}

We find that the PS1 AstrometryCorrection table does not fully cover the outskirts of our Draco field. So we select stars with a central rectangular region of $1.6\degree\times1.6\degree$ (six times the half-light radius) to make the density profile of Draco. Similarly, the candidate member stars within this region is divided into $100\times100$ grids, and the resulting stellar density map is smoothed using a gaussian kernel with a sigma of 2 pixels, as shown in \figref{ddensity}. The density map shows that Draco exhibits a density profile of typical dwarf elliptical galaxy, while no tidal features are detected above the $1\sigma$ noise level (\roughly0.04 in the figure). These results are consistent with those of \cite{Mu2018}, whose observations reach \roughly2\magn deeper than ours.

However, we notice a saddle-shaped density in the central contour, which is an artifact caused by saturated stars. In particular, a source with a \Gaia $G=8.68\magn$, located within the half-light radius of Draco, significantly affects the photometry of nearby stars. Saturation and bleed trails from this saturated star lead to the exclusion of nearby member stars from selection, resulting in an artificially low stellar density in that region. Furthermore, from visual inspection of the original images, we find that the saturation artifacts are more pronounced in the $g$ band than in the $r$ band. This conclusion is supported by the results shown in \figref{comp}, where the purity in the $g$ band is about 5\% lower than that in the $r$ band.

To estimate the structural parameters of Draco, we adopt the following four models, the Exponential model:
\begin{equation}
\begin{array}{lcl}
  \Sigma_{\rm exp}(r) & = & \Sigma_{\rm 0}{\rm exp}\left({-{r \over r_{E}}}\right), \\
\end{array}
\end{equation}
the Plummer model:
\begin{equation}
\begin{array}{lcl}
  \Sigma_{\rm Plummer}(r) & = & \Sigma_{0}\left(1+{r^{2} \over r_{P}^{2}}\right)^{-2}, \\
\end{array}
\end{equation}  
the King model:
\begin{equation}
\begin{array}{lcl}
  \Sigma_{\rm King}(r) & = & \Sigma_{0} \left[\left(1+ \frac{r^{2}}{r_{c}^{2}}\right)^{-\frac{1}{2}} - \left(1+ \frac{r_{t}^{2}}{r_{c}^{2}}\right)^{-\frac{1}{2}}\right]^{2}, \\
\end{array}
\end{equation}
and the S\'ersic model:
\begin{equation}
\begin{array}{lcl}
  \Sigma_{\rm Sersic}(r) & = & \Sigma_{0}{\rm exp}\left[{-\left({r \over r_e}\right)^{1/n}}\right]. \\
\end{array}
\end{equation}
In addition, the coordinate transformation is carried out according to the following equations:
\begin{equation}
\left\{
\begin{array}{l}
  r(x, y)^2 = A^2+(\frac{B}{1-\epsilon}) \\
  A = (x-x_0)cos(\theta)+(y-y_0)sin(\theta) \\
  B = (x-x_0)sin(\theta)+(y-y_0)cos(\theta) \\
\end{array}
\right.
\end{equation}
Here, $r_E$ and $r_P$ are the exponential and Plummer scale lengths, $r_c$ and $r_t$ are the King core and tidal radii, and $r_e$ is the effective radius in the Sérsic model. $\epsilon$ and $\theta$ denote the ellipticity and the position angle of the galaxy, respectively. We exclude regions affected by saturated stars and determine the structural parameters using the least-squares method with the \astropy \modelsd package. A summary of the fitted parameters is shown in \tabref{sp}. Our measurements of structural radii from different models are generally consistent with those reported by \cite{segal2007} and \cite{Mu2018}. The radii derived from the Plummer and King models are slightly larger (\roughly1\arcmin) than those in these previous studies, which may be attributed to their deeper (\roughly2\magn) observations. And our limiting magnitude only reaches the main-sequence turnoff (MSTO), resulting in smaller number of detected member stars and possible biases in the derived parameters.

\renewcommand{\arraystretch}{1.0}
\setlength{\tabcolsep}{2.5pt}
\begin{table}[h]
  \flushleft
  \caption{Structural parameters of Draco}
  \label{tab:sp}
  \begin{tabular}{lccccc}
  \hline
     Exponential & $\epsilon_{E}$ & $\theta_{E}$ & $r_E$ (\arcmin) &  &\\
  \hline
                 & $0.34\pm0.02$ & $88\pm2$ & $6.28\pm0.14$& &\\
                 &&&&&\\
     Plummer & $\epsilon_{P}$ & $\theta_{exp}$ & $r_P$ (\arcmin)&  &\\
                 & $0.34\pm0.02$ & $88\pm2$ & $10.89\pm0.22$& &\\
                 &&&&&\\
     King & $\epsilon_{K}$ & $\theta_{K}$ & $r_{c,K}$ (\arcmin)& $r_{t,K}$ (\arcmin)& \\
          & $0.34\pm0.02$ & $88\pm2$ & $8.31\pm0.35$& $39.49\pm2.34$&\\
    &&&&&\\
     S\'ersic & $\epsilon_{S}$ & $\theta_{S}$ & $r_{e,S}$ (\arcmin)& $n$ & \\
     & $0.34\pm0.02$ & $88\pm2$ & $9.53\pm0.21$& $0.75\pm0.03$&\\
  \hline
  \end{tabular}
\end{table}

\renewcommand{\arraystretch}{1.2}
\setlength{\tabcolsep}{5pt}
\begin{deluxetable*}{lccccccc}[t]
  \tablecaption{Positions and Luminosity Parameters \label{tab:property}}
  \tablehead{
  \colhead{ } & 
  \colhead{\ra} & 
  \colhead{\dec} & 
  \colhead{$(m-M)_0$} & 
  \colhead{$r_e$} & 
  \colhead{$M_g$} & 
  \colhead{$M_r$} & 
  \colhead{$M_V$} \\
  \colhead{} & 
  \colhead{deg} & 
  \colhead{deg} & 
  \colhead{\magn} & 
  \colhead{arcmin} & 
  \colhead{\magn} & 
  \colhead{\magn} & 
  \colhead{\magn}
  }
  \startdata
  Boötes III & 209.3 & 26.8 & 18.35~\tablenotemark{a} & --- & $-6.49\pm0.19$ & $-6.99\pm0.16$ & $-6.79\pm0.18$ \\
  Draco      & $260.053\pm0.001$ & $57.928\pm0.002$ & $19.557\pm0.026$ ~\tablenotemark{b}& $9.53\pm0.21$ & $-8.24\pm0.03$ & $-8.83\pm0.02$ & $-8.59\pm0.03$ \\
  \enddata
  \tablenotetext{a}{\scriptsize\cite{McConnachie2012}}
  \tablenotetext{b}{\scriptsize\cite{Bhardwaj2024}}
\end{deluxetable*}

\subsection{Boötes III and Styx Stellar Stream} 

The Boötes III and Styx stellar stream were simultaneously discovered by \cite{Grillmair2009}. The distance of the Styx stream from the Sun was estimated to be between 38 kpc and 50 kpc, and the Boötes III was identified as an overdensity embedded in this extended stream, which spans at least 53\degree. Given their similar distances and disturbed morphology of Boötes III, he argued that Boötes III may be the progenitor of Styx stream. \cite{Carlin2018} calculated the PM and orbit of Boötes III based on \Gaia DR2 PMs and radial velocities (RVs) derived from spectroscopic observations \citep{Carlin2009b}. Their PM measurements are consistent with the predicted retrograde orbit of Styx \citep{Grillmair2009}, which further supports the connection between Styx and Boötes III.

\begin{figure}[h]
  \center
  \includegraphics[width=\columnwidth]{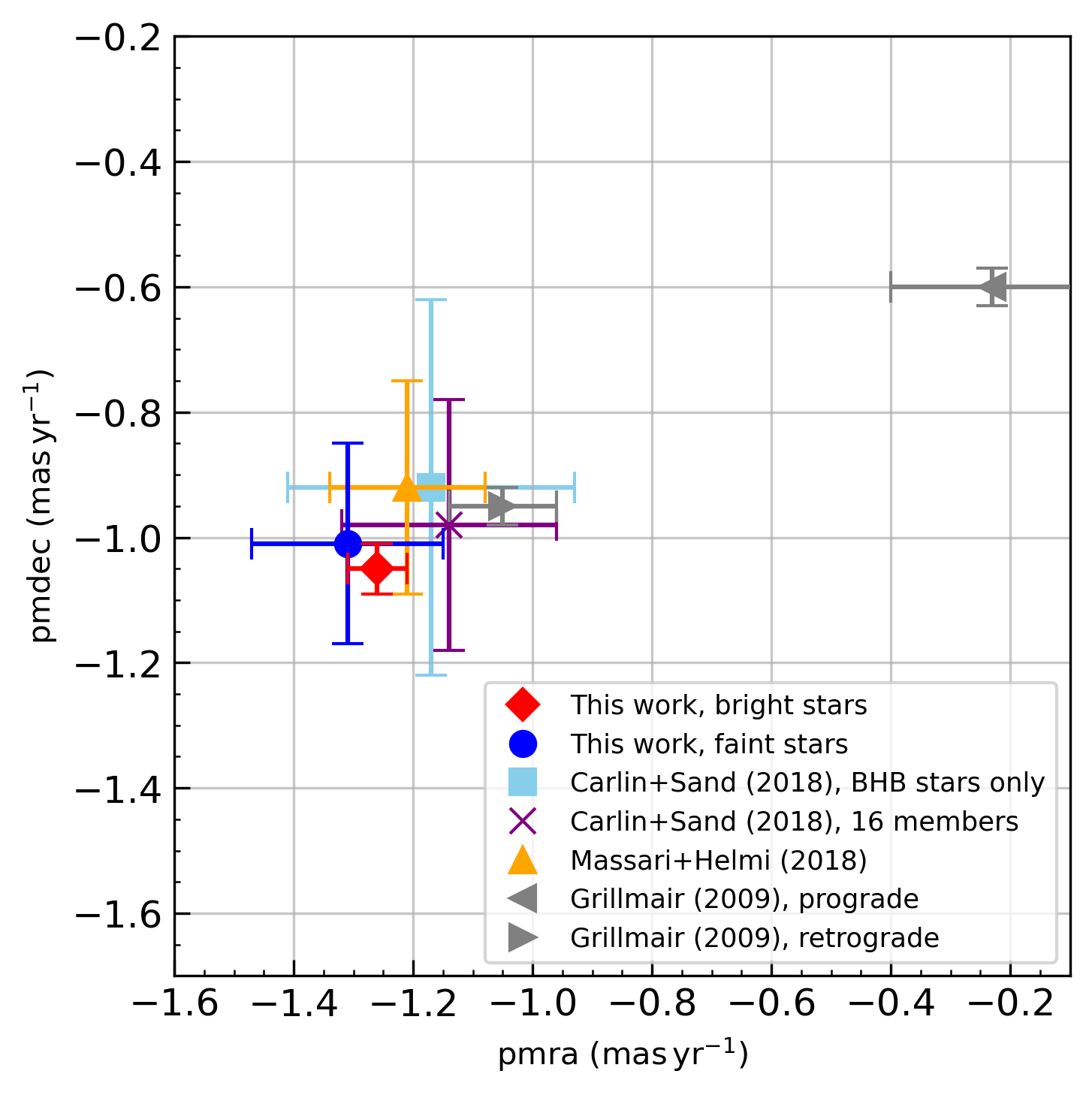}
  \caption{Measured PMs of Boötes III from different studies are represented by various symbols, as illustrated in the legend of this diagram. The sky-blue rectangle shows PM measurements based on six BHB member stars, while the estimation from 15 RV member stars plus one known RR Lyrae star is colored by purple \citep{Carlin2018}. The measurement from \cite{Massari2018} is coded by orange triangle. The predicted PMs based on orbit derivations from  \cite{Grillmair2009} are also shown in this figure.}
  \label{fig:pm_com}
\end{figure}

However, since the limiting magnitudes of PS1 and SDSS are near the MSTO of Boötes III, previous studies have poorly constrained its structural parameters, and the observed spatial distribution may be imcomplete due to the limited detection of member stars. The deeper observations from WFST resolves more faint member stars, allowing for a more detailed mapping of the spatial structure of Boötes III. By leveraging the high-precision PMs of bright Gaia DR3 stars and the PMs of fainter stars from our catalog, we derive a well-constrained PM for Boötes III. A comparison between our measured PMs and previous studies is shown in \figref{pm_com}. Our PM measurements are consistent with the predicted retrograde orbit of Styx. Furthermore, the morphology of Boötes III indicates that it has been tidally stretched along the direction of its PM. These results provide more direct evidence supporting the hypothesis of that Boötes III is the remnant of the progenitor galaxy that gave rise to the Styx stream.
   
It is worth noting that, according to measurements from \cite{Grillmair2009}, the Styx stream has a FWHM of 3.3\degree, indicating that our obsevation lies entirely within the stream and may include contributions from Styx member stars. However, as the Styx stream is much fainter than Boötes III, its influence on our results is limited. With the upcoming WFST WFS program, which will provide deeper observations and broader coverage of the Boötes III region, it will be possible to explore the morphological and kinematic connection between the Styx stream and Boötes III galaxy on a larger scale.

\section{Summary}
\label{sec:sum}
In this paper, we present the data reduction and the extraction of the final photometric catalog for the Boötes III and Draco satellite galaxies, with the data from the WFST pilot observing program. Additionally, we introduce a method to remove contamination from foreground stars and background galaxies. The main conclusions are summarized as follows:

\begin{enumerate}
  \item WFST shows a capability in measuring stellar PMs by combining data with PS1 DR2, achieving uncertainties of \roughly1.8\masy at 21\magn and \roughly3.0\masy at 22\magn in the $r$ band. Furthermore, the identification of a larger number of faint candidate member stars significantly improves the precision of PM measurement for these two galaxies. As a result, WFST achieves a PM measurement accuracy of \roughly0.2\masy based on stars fainter than $g=21\magn$.
  \item We measure the PMs for two satellite galaxies using the \Gaia PMs of bright stars: Boötes III has a PM of $\pmra=-1.26\pm0.05$\masy and $\pmdec=-1.05\pm0.04$\masy, while Draco has $\pmra=-0.05\pm0.01$\masy and $\pmdec=-0.24\pm0.01$\masy. Additionally, we derive the PMs using the faint candidate member stars by combining WFST data with PS1 DR2: Boötes III has $\pmra=-1.31\pm0.16$\masy and $\pmdec=-1.01\pm0.16$\masy, while Draco has $\pmra=-0.57\pm0.06$\masy and $\pmdec=-0.24\pm0.06$\masy. The consistency between these two measurements suggests that no significant variation in proper motion across stellar luminosity as these satellite galaxies undergoing the tidal interactions with the MW.
  \item Based on our density and contour maps, we find no evidence of tidal features in Draco. Structural parameters of Draco derived using exponential, Plummer, King, and Sérsic models are in good agreement with those obtained from previous and deeper observations.
  \item Based on the density map and PM measurements of Boötes III, we identify a high-density region in the central area where the PMs of bright stars are consists with those of faint stars. The morphology of this region appears to be shaped by the tidal forces of the MW, showing a significant elongation along the direction of its PM.
  \item By comparing our PM measurements of Boötes III with previous studies and the predicted retrograde orbit of Styx, we suggest a possible connection between these two stellar systems. The Styx stream may have originated from a satellite galaxy that was tidally disrupted by the Milky Way, with Boötes III representing the surviving core of this progenitor system.
\end{enumerate}

\section*{Acknowledgments}

This work is supported by the National Key Research and Development Program of China (2023YFA1608100), the National Science Foundation of China (12173088, 12233005, 12073078), the China Manned Space Program with grants with nos. CMS-CSST-2025-A20 and CMS-CSST-2025-A08, and STCSM through grant No. 24DX1400100. The Wide Field Survey Telescope (WFST) is a joint facility of the University of Science and Technology of China, Purple Mountain Observatory.

\bibliography{ms}{}
\bibliographystyle{aasjournal}

\end{document}